\documentclass{article}
\usepackage{amsmath,amsthm}
\usepackage{amssymb,latexsym}
\usepackage[mathscr]{eucal}
\usepackage{setspace}
\usepackage{graphics}
\usepackage{array}

\newcommand{\beq}{\begin{equation}}
\newcommand{\eq}{\end{equation}}

\newcommand{\ti}{\times}

\newcommand{\bfv}{{\bf v}}
\newcommand{\bfB}{{\bf B}}

\newcommand{\bfx}{{\bf x}}
\newcommand{\bfq}{{\bf q}}
\newcommand{\bfp}{{\bf P}}
\newcommand{\bfo}{{\bf\omega}}
\newcommand{\rp}{\right)}
\newcommand{\lp}{\left(}
\newcommand{\rb}{\right]}
\newcommand{\lb}{\left[}

\newcommand{\Sl}{\int_S}

\usepackage{graphicx}
\graphicspath{{converted_graphics/}}
\begin{document}

\title{\bf Divorticity and Dihelicity in Two-dimensional Hydrodynamics}         
\author{B.K. Shivamoggi\footnote{\large {Permanent Address: University of Central Florida, Orlando, FL 32816-1364}} and G.J.F. van Heijst\\
J.M. Burgers Centre and Fluid Dynamics Laboratory\\Department of Physics\\
Eindhoven University of Technology\\
NL-5600MB Eindhoven, The Netherlands\\
\\
J. Juul Rasmussen\\Optics and Plasma Research Department\\OPL-128, Riso National Laboratory\\Technical University of Denmark\\P.O. Box 49\\DK-4000 Roskilde, Denmark}        
\date{}
\maketitle

\large\section{\bf Abstract}

\indent 

A framework is developed based on the concepts of 
{\it divorticity} ${\textbf B}$($\equiv\nabla\times\bfo$, $\bfo$ being the vorticity) and $\textit{dihelicity}~ g \lp \equiv\bfv\cdot\textbf{B}\rp$ for discussing the theoretical structure underlying 
two-dimensional (2D) hydrodynamics. This formulation leads to the global and Lagrange invariants that could impose significant constraints on the evolution of divorticity lines in 2D hydrodynamics.

\section{Introduction}
\indent

The theory of two-dimensional (2D) fully developed turbulence (FDT) (Kraichnan \cite{Kra}, Batchelor \cite{Bat}), until recently, remained almost an academic exercise, not withstanding its possible connections with large-scale flows in thin fluid shells such as atmospheres and oceans. 2D FDT has now been produced to a close approximation in a variety of laboratory experiments (Couder \cite{Cou}, Kellay et al. \cite{Kell}, Martin et al. \cite{Mar}, Rutgers \cite{Rut}, Rivera et al. \cite{Riv}, Vorobieff et al. \cite{Voro}, Rivera et al. \cite{Riv1}, \cite{Riv2}) (see Clercx and van Heijst \cite{Cle} for a recent review as well as references).

\indent Vorticity at very small length scales behaves like a passive scalar (Weiss \cite{Wei}, Falkovich and Lebedev \cite{Fal}, and Nam et al. \cite{Nam}) and is advected by the large-scale flow structures (Legras te al. \cite{Leg}, Chen et al. \cite{Che}). This leads to thin sheets with large vorticity gradients (Saffman \cite{Saff}). Kuznetsov et al. \cite{kuzn} gave qualitative arguments to support the formation of sharp vorticity gradients in 2D hydrodynamics even for smooth initial conditions. The $\textit{divorticity}~ \textbf{B} \lp\equiv\nabla\times\bfo,~ \bfo ~\text{being the vorticity}\rp$ (Kuznetsov et al. \cite{kuzn}, Kida  \cite{Kid}) amplification provides the physical mechanism underlying the enstrophy cascade in 2D. (This is like the vortex stretching process underlying the energy cascade in 3D.) Indeed, Bruneau et al. \cite{Bru} identified filamentary structures occurring in highly strained regions as those being responsible for the enstrophy cascade. Further, there is selective rapid decay of vorticity in these layers because such regions experience typically stronger viscous diffusion than other regions. Consequently, divorticity sheets are more likely to occur near vorticity nulls \footnote{\large Vorticity nulls are seats of vortex reconnection in 3D (Kuznetsov \cite{Kuz2}).} (Shivamoggi \cite{Shi}), just as vortex sheets form near velocity nulls in 3D. It therefore appears to be useful to develop a framework  based on the concepts of divorticity $\textbf{B}$ and \textit{dihelicity} $g\lp \equiv \bfv\cdot \textbf{B} \rp$ for discussing the theoretical structure underlying 2D hydrodynamics, which is the objective of this paper.

\section{Divorticity Evolution in 2D}

\indent The equation of motion in 2D ideal hydrodynamics (in the usual notation) is

\beq\tag{1a}
\frac{D \bfv}{D t} \equiv \frac{\partial{\bf v}}{\partial t} +\lp\bfv\cdot\nabla\rp\bfv=-\frac{1}{\rho} \nabla p
\eq

\noindent or

\beq\tag{1b}
\frac{\partial{\bf v}}{\partial t}- \bfv\times\bfo= \nabla \lp\frac{p}{\rho}+\frac{1}{2}\bfv^2\rp 
\eq

\indent
On taking the curl of equation (1), we obtain the vorticity evolution equation  

\beq\tag{2}
\frac{\partial{\bfo}}{\partial t}=\nabla\times\lp\bfv\times\bfo\rp
\eq

\noindent
which in 2D leads to , 

\beq\tag{3a}
\frac{D \bfo}{D t}=0.
\eq

\noindent or

\beq\tag{3b}
\bfo =const,
\eq

\noindent
so the vorticity $\bfo$ becomes irrelevant in 2D.\footnote{\large In 3D, on the contrary, the vorticity becomes invariant only with respect to the motion of material particles along the vortex lines in the Lagrangian picture (Kuznetsov and Ruban \cite{Kuz3}). This property is exploited in developing the so called \textit{vortex line representation} to provide a convenient framework to analyze the vortex line breaking process underlying the finite-time singularity development in 3D (\cite{Kuz3}).}

\indent On the other hand, introducing the divorticity $\textbf{B}$ (Kuznetsov et al. \cite{kuzn}, Kida \cite{Kid}) 

\beq\tag{4}
\bfB\equiv\nabla\times\bfo
\eq

\noindent we obtain from equation (2) in 2D (Kuznetsov et al. \cite{kuzn}) the divorticity evolution equation  

\beq\tag{5a}
\frac{\partial \bfB}{\partial t}=\nabla\times\lp\bfv\ti\bfB\rp
\eq

\noindent or
\beq\tag{5b}
\frac{D \bfB}{D t}=\lp\bfB\cdot\nabla\rp\bfv
\eq

\noindent
which is very much like the vorticity evolution equation in 3D.

\section{Global Dihelicity Invariant}

\indent
Suppose $C$ be a closed curve enclosing an area $S$ and moving with the fluid. Consider the total \textit{dihelicity} $G$  
 
\beq\tag{6}
G \equiv \Sl\bfv\cdot\bfB~ dA.
\eq

\noindent
Then, we have the following result:

\noindent\textbf{Thereom}: The total dihelicity is invariant in 2D ideal hydrodynamics.

Proof: We have, on using equations (1) and (5), 

\beq\tag{7}
\begin{array}{lcr}
\displaystyle\frac{D}{D t}\lp\bfv\cdot\bfB\rp=\bfB \cdot\frac{D \bfv}{D t} + \bfv\cdot\frac{D \bfB}{D t}\\
\\
\displaystyle~~~~~~~~~~~~~= -\frac{\bfB}{\rho}\cdot\nabla p + \bfv\cdot\lp\bfB\cdot\nabla\rp\bfv\\
\\
\displaystyle~~~~~~~~~~~~~=\bfB\cdot\nabla\lp-\frac{p}{\rho}+\frac{1}{2}\bfv^2\rp.
\end{array}
\eq

\indent
Using equation (7), we then obtain

\beq\tag{8}
\begin{array}{ccl}
\displaystyle\frac{DG}{D t}&=&\displaystyle\Sl\frac{D}{D t}\lp\bfv\cdot\bfB\rp dA\\
\\
&=& \Sl\bfB\cdot\nabla\lp-\displaystyle\frac{p}{\rho}+\frac{1}{2}\bfv^2\rp dA\\
\\
&=&\Sl\nabla\cdot\lb\bfB\lp-\displaystyle\frac{p}{\rho}+\frac{1}{2}\bfv^2\rp\rb dA\\
\\
&=&\oint_c\lb-B_y\lp-\displaystyle\frac{p}{\rho}+\frac{1}{2}\right.\bfv^2\rp dx+ B_x\left.\lp-\displaystyle\frac{p}{\rho}+\frac{1}{2}\bfv^2\rp dy\rb\\
\\
\displaystyle&=&0
\end{array}
\eq

\noindent on imposing the boundary condition $\hat{\textbf{n}}\cdot\bfB=0$ on $C$ ($\hat{\textbf{n}}$ being the outward normal to $C$). Thus, we have in 2D ideal hydrodynamics the total \textit{dihelicity} invariant\footnote{\large It may be noted (the authors are thankful to Dr. David Montgomery for this observation) that (6) may be rewritten as

\beq
\begin{aligned} 
G&=\int_S \bfv\cdot\nabla\times\bfo~ dA\\
&=\int_S\lb\nabla\cdot\lp\bfv\times\bfo\rp+\omega^2\rb dA\\
&=\oint_C\lb\lp\bfv\times\bfo\rp_x dy-\lp\bfv\times\bfo\rp_y dx\rb+\int_S\bfo^2 dA.
\end{aligned}
\eq
For the case of a periodic box, the line integral above vanishes and $G$ degenerates into the usual global enstrophy invariant.}

\beq\tag{9}
G=const.
\eq

This invariant could impose a significant constraint on the evolution of divorticity lines in 2D hydrodynamics.

\section{Impulse Formulation}

\indent~~ Impulse formulations of 3D hydrodynamic equations were considered by Ku\'{z}min \cite{Kuz}. This led to the result, upon the use of an appropriate gauge condition, that the helicity, which is important for the study of topological properties of vorticity lines (Moffatt \cite{Moff}), is a Lagrange invariant. Let us now proceed to develop impulse formulations of 2D hydrodynamic equations.

\indent ~~Put,

\beq\tag{10}
\bfq = \bfv + \nabla\phi,
\eq

\noindent where $\phi$ is chosen so that the impulse velocity $\bfq$ has a compact support. (10) also constitutes a Weber's transform (Lamb \cite{Lam}), as mentioned by Kuznetsov \cite{Kuz2}.

\indent~~~~~ Note that fluid impulse (Batchelor \cite{Bat1}) is then given by

\beq\tag{11}
\begin{array}{ccl}
\bfp &=&\displaystyle\frac{1}{2}\int\bfx\ti\bfo~ dA\\
\\
&=& \displaystyle\frac{1}{2}\int\bfx\ti\lp\nabla\ti\bfq\rp dA\\
\\
&=&\displaystyle\frac{1}{2}\int\bfq ~dA\\
\end{array}
\eq

\noindent so that $\displaystyle\frac{1}{2}~ \bfq$ is the impulse density for 2D hydrodynamics.

Then, we obtain from equation (1), 

\beq\tag{12}
\frac{\partial \bfq}{\partial t}-\bfv\ti\lp\nabla\ti\bfq\rp=-\nabla\lp\frac{p}{\rho}+\frac{1}{2}\bfv^2-\frac{\partial \phi}{\partial t}\rp.
\eq

\indent On imposing the following gauge condition on $\phi$,  

\beq\tag{13}
\frac{\partial \phi}{\partial t}+\lp\bfv\cdot\nabla\rp\phi+\frac{1}{2}\bfv^2-\frac{p}{\rho}=0
\eq

\noindent equation (11) becomes

\beq\tag{14a}
\frac{D \bfq}{D t}=-\lp\nabla\bfv\rp^T\bfq
\eq

\noindent or

\beq\tag{14b}
\frac{\partial \bfq}{\partial t}=-\nabla\lp\bfq\cdot\bfv\rp.
\eq

\section{Dihelicity Lagrange Invariant and Beltrami States}

\indent On combining equations (5b) and (14a), we obtain

\beq\tag{15}
\frac{D}{D t}\lp\bfq\cdot\bfB\rp=0
\eq

\noindent which leads to the \textit{dihelicity} Lagrange invariant   

\beq\tag{16}
\bfq\cdot\bfB=const.
\eq

On the other hand, equation (14b) yields in the steady state  

\beq\tag{17}
\bfq\cdot\bfv=const.
\eq

Combining (16) and (17), we obtain for the 2D Beltrami state  

\beq\tag{18}
\bfB=a \bfv
\eq

\noindent $a$ being an arbitrary constant, which is totally consistent with equation (5a).

\indent In order to gain physical insight into the \textit{dihelicity} Lagrange invariant (16), note that if ${\bf \ell}$ is a vector field associated with an infinitesimal line element of the fluid, ${\bf\ell}$ evolves according to (Batchelor \cite{Bat1})

\beq\tag{19}
\frac{D \mathbf{\ell}}{D t}= \lp\mathbf{\ell}\cdot\nabla\rp\bfv
\eq

\noindent which is identical to the equation of evolution of divorticity, namely, equation (5b). Therefore, the divorticity lines evolve as line elements. Thus, the $\textit{dihelicity}$ Lagrange invariant (16) in 2D hydrodynamics physically signifies the constancy of the action in $\bfq$-space. It may be noted that the helicity Lagrange invariant in 3D hydrodynamics deduced by Ku\'{z}min \cite{Kuz} also physically signifies the constancy of the action in $\bfq$-space. In this sense, the \textit{dihelicity} Lagrange invariant appears to be the exact 2D analogue of the helicity Lagrange invariant for the 3D case.

\section{Discussion}
\noindent The vorticity advection at small scales by the large-scale flow structures, according to general belief, leads to thin sheets with large vorticity gradients. So, the divorticity amplification appears to provide the physical mechanism underlying the enstrophy cascade in 2D somewhat like the vortex stretching process underlying the enstrophy cascade in 3D. It therefore appears to be useful, as we saw in the foregoing, to develop a framework based on the concepts of divorticity $\bfB$ and \textit{dihelicity} $g\equiv\bfv\cdot\bfB$ for discusssing the theoretical structure underlying 2D hydrodynamics. This formulation leads to global and Lagrange invariants that could impose significant constraints on the evolution of divorticity lines in 2D hydrodynamics.  

\section {Acknowledgements}

\indent The authors are thankful to Dr. David Montgomery for his helpful comments. BKS would like to thank The Netherlands Organization for Scientific Research (NWO) for the financial support.

\end{document}